\documentstyle[11pt,newpasp,twoside,epsf]{article}
\def\edcomment#1{\iffalse\marginpar{\raggedright\sl#1\/}\else\relax\fi}
\reversemarginpar

\begin{document}
\title{Gas Properties in the Starburst Centers of Barred Galaxies}
 \author{S.\ H\"uttemeister}
\affil{Astronomical Institute, Ruhr-Universit\"at Bochum, 
Universit\"atsstra\ss{}e 150, 44780 Bochum, Germany}
\author{S.\ Aalto}
\affil{Onsala Space Observatory, 43992 Onsala, Sweden}

\begin{abstract}
The physical properties of molecular gas can be deduced from 
the analysis of line intensity ratios, most commonly of the CO 
isotopomers $^{12}$CO and  $^{13}$CO. Barred galaxies, especially those
with central starbursts, provide an excellent laboratory to study the
changing properties of the gas as it approaches the center. Four galaxies
(UGC\,2855, UGC\,2866, NGC\,7479 and NGC\,4123) are presented as examples
of the results that can be obtained. Special attention is given to the 
case of (abnormally) high $^{12}$CO/$^{13}$CO (${\cal R}_{12/13}$) line 
ratios (exceeding 20), which may indicate unusual gas properties. 
Qualitative scenarios for the structure of the ISM explaining a range
of values of ${\cal R}_{12/13}$ are discussed.
\end{abstract}

\section{Gas flows in bars and changing gas properties}
The importance of gas flows in bars as a transfer mechanism of material
toward the centers of galaxies, thus fuelling central activity (certainly
starburst and possibly AGN-related) and driving galaxy evolution is a
well-established fact (e.g.\ Combes, Dupraz, \& Gerin 1990 and references
therein, Sakamoto et al.\ 1999). Gas flowing along the bar loses angular
momentum, resulting in a net infall that can be as large as several solar 
masses per year.  Stable, non-intersecting orbits for gas are required; 
between corotation and the inner Lindblad resonance(s) (ILRs) the gas 
moves on elongated, bar-sustaining $x_1$-orbits. Material may accumulate 
close to the ILR(s), a `spray' of gas may impact gas still on  $x_1$-orbits, 
and characteristic, curved bar shocks, often accompanied by dust lanes, 
may develop. 

Thus, it is easy to see how a {\em diffuse, unbound} gas component may
form along the bar, by either tidal disruption of bound clouds or cloud
collisons (Das \& Jog 1995, H\"uttemeister et al.\ 2000). In time, as the
bar evolves, the bulk of the gas is funneled from the outer $x_1$-orbits 
to an inner family of (slightly anti-bar) $x_2$-orbits (e.g.\ Friedli \& 
Benz 1993). Strength, shape and existence of the bar shock depend on the 
central mass concentration (see e.g.\ simulations by Athanassoula 1992). 

When the gas approaches the central region, part of it may remain diffuse
and unbound. However, it may also pile up and become dense enough to 
cause (and then continue to feed) a nuclear starburst. Clearly, the gas
properties within the bar and within the central starburst region 
should differ. The gas {\em within} the starburst center is also expected
to consist of a number of different phases, e.g.\ partly diffuse material
flowing in from the bar, dense clouds that may represent both hot, actively
star forming cores and a dense component that does not (at the moment) 
form stars and thus cools efficiently, gas exposed to strong shocks and --
as the starburst evolves -- a component that is strongly affected by the 
energy released by the young massive stars and thus disrupted. The balance
of these phases will change with the state of the burst; therefore the
gas properties, if they can be derived accurately enough, can be used 
as an indicator of starburst evolution. 

{\em Line intensity ratios}, especially of the $^{12}$CO and $^{13}$CO
isotopomers, which are sensitive to gas densities of a few 100\,cm$^{-3}$, 
but also, if available, of high density tracers like HCN, HNC
and CN, are powerful diagnostics of the physical properties of the gas 
they arise from. This requires observations that are still at the limit
of what the current generation of mm-telescopes (both single dish and
interferometers) can do, since it is essential to observe both higher 
$J$ transitions (at least the 2--1 line) and rare isotopomers. The data
obtained are then analysed by non-LTE radiative transfer models.  

\section{Sample galaxies}

\begin{table}[t]
\caption{Summary of the line ratios observed in the four sample
galaxies \vspace{2mm}}
\begin{tabular}{r|l|l|l}
Galaxy &  ${\cal R}_{12/13}$(1--0) & ${\cal R}_{12/13}$(2--1) & 
$^{12}$CO/HCN(1--0) \\
\hline
UGC\,2855 (center) & $\sim 10$ & $\sim 14$ & -- \\
\hline
UGC\,2866 (center) & $\sim 25$ & $\sim 10$ & $> 38$ \\
\hline
NGC\,7479 (global) & $\sim 30$ & -- & -- \\
(center)           & 23 (10 -- 30)$^a$ & 13 & $\sim 25$ \\
(along bar)        & 4 -- $> 40$ & -- & -- \\
\hline
NGC\,4123 (center) & 26 & 14 & $> 32$ \\
(bar end)          & 4  &  4 & -- \\
\end{tabular} \vspace{2mm}  \\
$a)$: For NGC\,7479, 10 -- 30 is the range encountered in high resolution
interferometric observations, while 23 is the average ratio over the 
central 1\,kpc radius.
\end{table}

\begin{figure}
\plotone{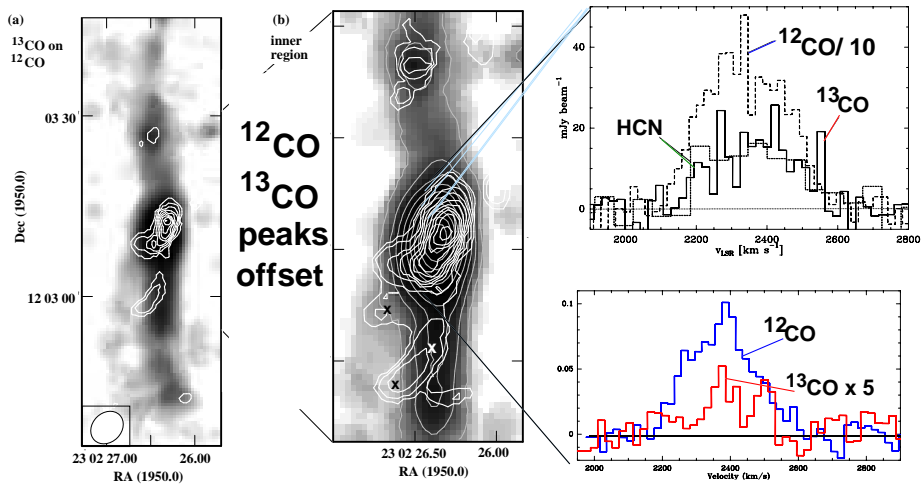}
\caption{The molecular distribution in the $^{12}$CO and $^{13}$CO 
(1--0) transitions in the barred starburst galaxy NGC\,7479 ($^{12}$CO:
greyscale, $^{13}$CO: contours). The right panels (a, b) show the central line
profiles; the upper panel displays the (1--0) transitions, measured
interferometrically with OVRO, while the lower panel shows the emission 
in the (2--1) lines, observed at a single dish telescope, the HHT.}
\end{figure}

We will now discuss the results of line intensity ratio studies done toward
four sample galaxies that show the range of ratios we can expect to
encounter. To relate the ratios discussed, a typical value for 
${\cal R}_{12/13}$(1--0) in the Galactic disk is $\sim 6$ (Polk et al.\
1988), (starburst) nuclei typically have ${\cal R}_{12/13}$(1--0) of 10 -- 15
(Aalto et al.\ 1995), and a few luminous infrared mergers show `abnormally
high' ratios of $> 20$ (Aalto et al.\ 1991, Casoli, Dupraz, \& Combes 1992).
The line ratios we have obtained for our sample galaxies are summarized in
Table\,1. 

{\bf UGC\,2855 and UGC\,2866:}
UGC\,2855 is a SBc spiral at a distance of $\sim$ 20\,Mpc, a weak 
starburst. UGC\,2866, its smaller companion, also has a bar morphology
and is a strong starburst with warm dust (IRAS $60\mu$m$/100\mu$m ratio
$\sim 0.8$). We have observed both galaxies with the OVRO interferometer
as well as with the OSO and HHT single dish telescopes (H\"uttemeister,
Aalto, \& Wall 1999). UGC\,2855, which we believe to be in a `pre-starburst'
state without strong bar shocks, still in the process of accumulating gas 
in the center, is characterized by `normal' values for ${\cal R}_{12/13}$(1--0)
and ${\cal R}_{12/13}$(2--1) (see Table\,1). 

In contrast, UGC\,2866 has a very high central
${\cal R}_{12/13}$(1--0) ($\sim 25$), a value so far only encountered
in a few starburst mergers. UGC\,2866, while a strong starburst, is not
a merger, and its FIR luminosity is moderate at $4.9 \cdot 10^{10}\,L_{\sun}$.
The value for ${\cal R}_{12/13}$(2--1) drops into the normal range ($\sim 10$).
The high $^{12}$CO/HCN(1--0) ratio of $> 38$ (only a limit for the HCN emission
could be obtained) is also noteworthy and interesting in a starburst galaxy
that should contain a significant amount of dense gas. 

{\bf NGC\,7479} is a well-studied SBc starburst spiral at $D = 32$\,Mpc (e.g.
H\"uttemeister et al.\ 2000, Laine et al.\ 1999). Globally, i.e.\ avaraged
over bar and center, ${\cal R}_{12/13}$(1--0) is very high, interferometric
observations of the central starburst region show significant variations
between normal and high ratios, with a `high' avarage of 23. Within the 
bar the variation is even more extreme, between extremely high ratios 
exceeding 40 and disk-like low ratios of $< 5$. Again, ${\cal R}_{12/13}$(2--1)
in the starburst center is within the normal range (13) (Fig.\,1). 

\begin{figure}[t!]
\plotone{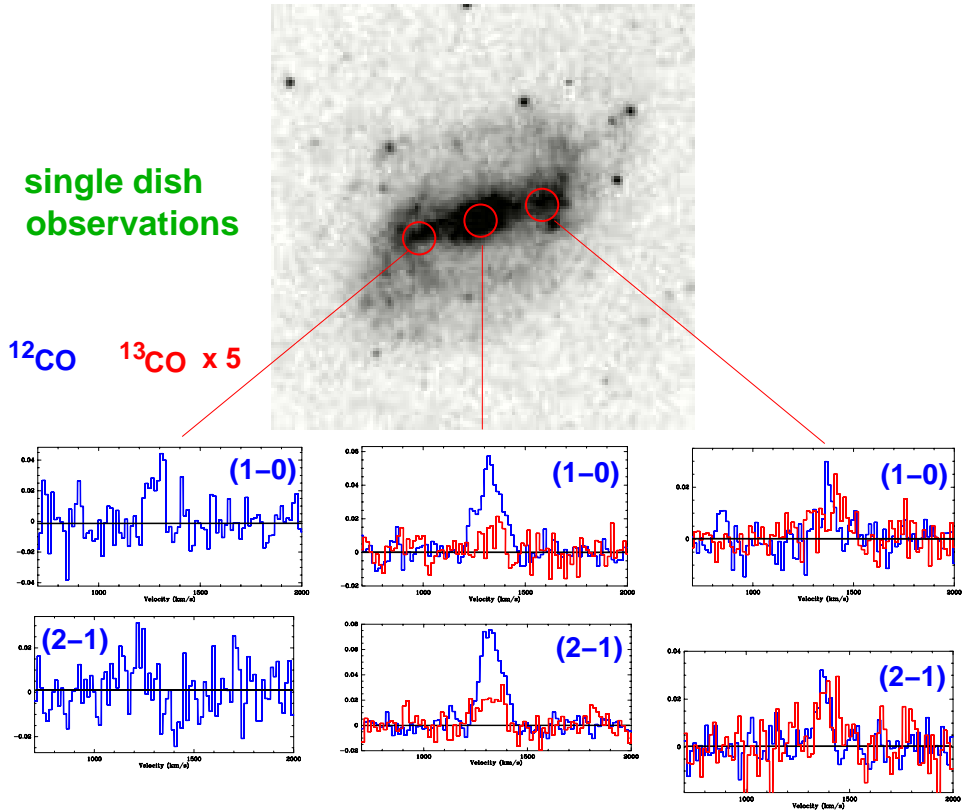}
\caption{$^{12}$CO and $^{13}$CO (1--0) and (2--1) emisison in the
barred starburst galaxy NGC\,4123 at the indicated positions in the central
region and toward the bar ends, observed at SEST. Note the extreme drop 
in ${\cal R}_{12/13}$ between center and bar end.}
\end{figure}

{\bf NGC\,4123} is a fairly inconspicuous SBc starburst galaxy with cool dust
(IRAS $60\mu$m$/100\mu$m ratio $\sim 0.6$). Still, once more we find 
a high ${\cal R}_{12/13}$(1--0) value of 26, accompanied by a normal
${\cal R}_{12/13}$(2--1) of 14 and undetected HCN(1--0) emission 
($^{12}$CO/HCN(1--0) $> 32$) (Fig.\,2). The very low (disk-like)
${\cal R}_{12/13}$ of $\sim 4$ at the bar ends is noteworthy. 

\section{High values for ${\cal R}_{12/13}$(1--0) and scenarios for 
cloud properties}

\begin{figure}
\plotone{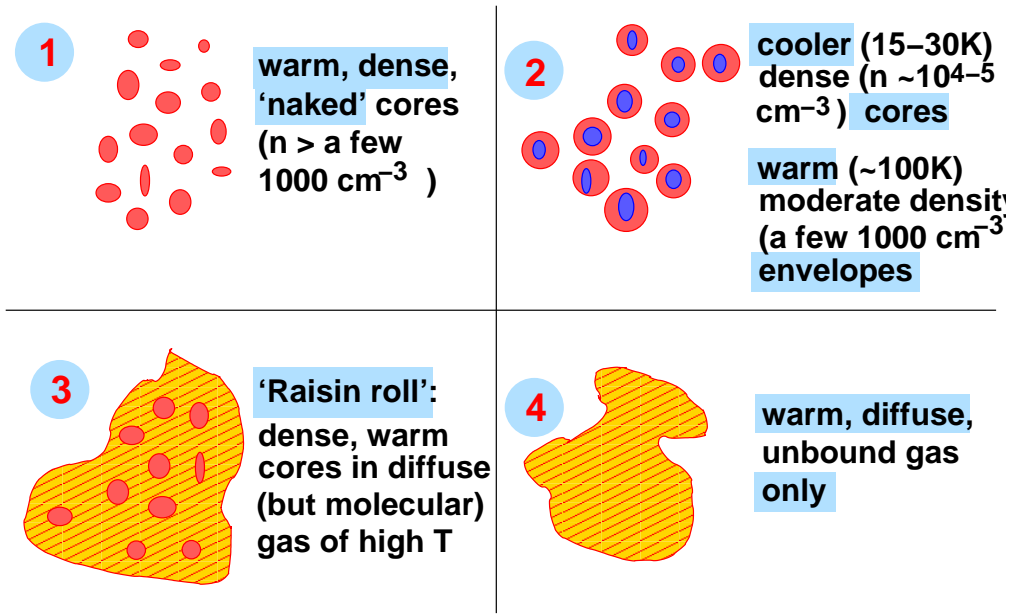}
\caption{Cartoon of possible scenarios of dominant cloud structure 
that may be encountered in a (starburst) nucleus, causing different 
values of ${\cal R}_{12/13}$ as discussed in the text. }
\end{figure}

The observations of the four sample galaxies described above leave us
with a somewhat surprising result: There seems to be a class of 
non-merger barred spiral (starburst) galaxies with moderate IR luminosities
that have a high ${\cal R}_{12/13}$(1--0), thought to be the exclusive 
property of a few extreme starburst mergers. While only a minority 
among the barred spirals as well as the mergers a `high-ratio' galaxies
(as will be further discussed  in a forthcoming paper examining a larger 
sample of starburst galaxies, H\"uttemeister \& Aalto, in prep.), it 
indicates variations of the dominating component of the molecular ISM 
within the same morphological group (barred galaxies and mergers). 
Thus, we can hope that with more detailed modelling we will be able to 
link these differences in  ${\cal R}_{12/13}$ and other line ratios to 
the evolutionary processes. 

Even now, we can examine the standard explanations for high 
${\cal R}_{12/13}$ put forward for mergers and see whether they are 
applicable in a barred starburst environment. Infall of low 
$^{13}$CO-abundance gas can be excluded for barred starbursts, since typically
${\cal R}_{12/13}$ drops toward the bar ends (see the example of NGC\,4123).
In situ nucleosynthesis effects in the central starburst also seem unlikely,
since high ratios are found as well along the bar. The selective in situ
removal of $^{13}$CO (e.g.\ by selective photodissociation) is possible
in principle and may play a role, especially in diffuse gas.
However, we typically find that ${\cal R}_{12/13}$(2--1) is normal when 
${\cal R}_{12/13}$(1--0) is high.
This may also be the case for most starburst mergers (though the opposite
has been claimed by e.g.\ Taniguchi, Ohyama \& Sanders 1999) and will be
discussed in our forthcoming paper. This behaviour leaves {\em excitation
effects} as the only reasonable explanation.

A simple one-component non-LTE radiative transfer model indicates high 
temperatures ($\sim 100$\,K) and moderate densities of a few $10^3$\,cm$^{-3}$
as a possible set of physical properties that cause a combination of
high values of ${\cal R}_{12/13}$(1--0) and moderate ${\cal R}_{12/13}$(2--1).
However, as pointed out in Sect.\,1, the molecular ISM in a starburst 
nucleus certainly consists of multiple components. In Fig.\,3, we show a 
cartoon of different simplified scenarios that may dominate in a
starburst nucleus. 

Panel\,1 depicts the one-component scenario described above. Panel\,2
shows a situation that may be realized in many starburst nuclei displaying
`normal' values of 10 -- 15 in both ${\cal R}_{12/13}$(1--0) and (2--1).
Clearly, some hot, dense cores are also needed to make a starburst,
but it is possible that the dense component cools so efficiently that a 
significant fraction of the material remains at temperatures of 15 -- 30\,K.
In this case, most $^{12}$CO emission arises in the envelopes, while more
than half of the $^{13}$CO emission originates in the cores. 
This scenario is probably realized in NGC\,1808 (Aalto et al.\ 1994).
In panel\,3, we display a `raisin roll' scenario, with warm cores embedded 
in a warm, unbound, diffuse intercloud medium. Here, almost all $^{12}$CO
emission arises in the diffuse medium, but $^{13}$CO originates almost 
exclusively in the cores. A medium like this may be encountered in the
high pressure environment of extreme starbursts and can give higher 
${\cal R}_{12/13}$(1--0) than (2--1) (Aalto et al.\ 1995). 
Finally, the situation described in
panel\,4, a purely diffuse medium, is unlikely to be realized in starburst
nuclei, but may be encountered along a molecular bar. Here, both 
${\cal R}_{12/13}$(1--0) and (2--1) are expected to be high. Clearly, other
combinations of the components shown are possible, and in more detailed 
modelling complementary information, e.g.\ on high density
tracers and possible positional offsets between the distribution of
different species have to be taken into account.

\section{Conclusions}
We have shown that the gas properties of the starburst centers of barred
galaxies can be analyzed by means of the line ratios of CO isotopomers.
Detailed modelling of the the significant variations encountered can
indicate the evolutionary state of the burst. High values of 
${\cal R}_{12/13}$(1--0) found in three sample galaxies are of special 
interest, since so far these properties were exclusively linked to 
probably shortlived, extreme burst phases of IR-luminous mergers containing
hot dust. We have presented simple scenarios that can explain variations
in ${\cal R}_{12/13}$, all based on changes in molecular excitation.

\nopagebreak

\end{document}